# Inverted anisotropy in the in-plane dissipation of BSCCO tapes: preliminary results

A. S. García-Gordillo, D. Antunez and E. Altshuler*

*Superconductivity Laboratory and Group of Complex Systems and Statistical Physics,*
*IMRE-Physics Faculty, University of Havana, 10400 Havana, Cuba*

J. Jiménez

*LUCES, IMRE, University of Havana, 10400 Havana, Cuba*

C. F. Sánchez-Valdés

*División de Materiales Avanzados, (IPICyT), México*

(Dated: May 25, 2018)

The electrical properties of superconducting tapes and coatings in the direction transverse to the long dimension of the composite has been rarely studied. However, transverse dissipation can eventually determine the behavior of a transmission line in the case of failure due to the presence of transversal cracks, and is also fundamental in the AC regime. In this paper we present a preliminary experimental study of the electrical transport properties along the transverse direction of BSCCO-metal tapes, and compare them with those measured along the long axis of the material. In spite of the fact that the tapes under study are not multi-filamentary, our experiments suggest that there is a measurable anisotropy of the transport properties between the longitudinal and transverse directions.

## I. INTRODUCTION

Inhomogeneities have been a major concern in developing real-world applications for superconducting materials. From bulk materials [1–7] to tapes [8, 9], coated conductors [10, 11] and thin films [12–14], the effects of inhomogeneities on the superconducting properties have been systematically studied –typically exploring the relations between structural morphology and the thermal, electric and magnetic properties.

For tapes and coated conductors, the presence of extended defects, and their effects on the supercurrent in the longitudinal direction, has been well documented [15–18]. In multifilamentary tapes, it is found that such defects do not affect much the inter-filamentary currents, which provide an "escape" route in case of interruption of the longitudinal current. Inter-filament currents are also fundamental for AC losses [19, 20]. However, the transverse transport has been little studied in composite superconductors: most work aim at the transport in the longitudinal direction.

The transverse critical current in samples of an Ag-sheathed multifilamentary $Bi_2Sr_2Ca_2Cu_3O_{10+x}$ (Bi-2223/Ag) tapes was studied in the 2000's using magneto-optical imaging (MOI)[21–23], showing a modest degree of anisotropy between the longitudinal and transverse dissipations only at low temperatures.

Later on, a clear anisotropy in the dissipation associated to *transport* currents in the longitudinal *vs.* transverse directions have been found and studied in detail by our group by cutting longitudinal and transversal bridges using a laser technique [25–28].

While the observed anisotropy could be expected in a multi-filamentary, it seems unlikely that it could be found in composites where the superconducting phase consists in a single layer. In this paper, we tackle that problem for the first time in the literature, as far as we know. From the measurements performed to the transverse direction, it was found that the values of critical current densities were different than those measured in the longitudinal direction, showing a clear anisotropy in the transport properties. The anisotropy found shows an anomalous behavior due to the fact that the values of critical current densities are larger in the transverse direction.

## II. EXPERIMENTAL

Samples were thin bridges cut from a superconducting tape in the longitudinal and transverse directions (Fig.1). Due to its commercial character, the knowledge about the composition or morphology of the original tape was insufficient for our purposes. In order to get more information on the tape, it was necessary to perform optical microscopy and X ray fluorescence studies.

---

* corresponding author, ealtshuler@fisica.uh.cu

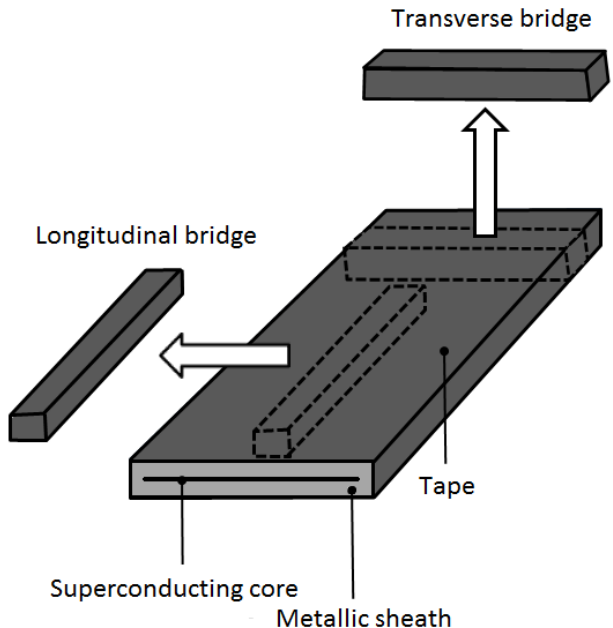

FIG. 1: Sketch showing the longitudinal and transverse bridges cut from the tape.

The tape was 4.4 mm wide and 0.5 mm thick. From the micrograph of the cross-section of the tape displayed in Fig.2 it is easy to see that we are in the presence of a non multi-filamentary tape. The superconductor forms a layer in the center of the tape and a material resembling copper covers it. After a semi-cuantitative study of the chemical elements of the tape using X-Ray spectroscopy, it was found that the outer part which covers the superconductor was made of copper, zinc and tin. This allows us to deduce that we are dealing with an alloy of these materials. Also bismuth and lead were found in the center of the tape, which strongly suggests that the superconducting phase corresponds to a system of BSCCO doped with lead. Summarizing, the sample contains BSCCO-Pb in the core of the tape and it is sheathed by an alloy of Cu,Zn and Sn. Transport measurements were performed along the longitudinal and transverse bridges using the four-probe technique. $I-V$ curves were obtained by increasing DC currents provided by a Philips PE 1540 DC power supply, while the voltage was measured by a virtual multimeter programmed in a PCI 6115 National Instruments card.

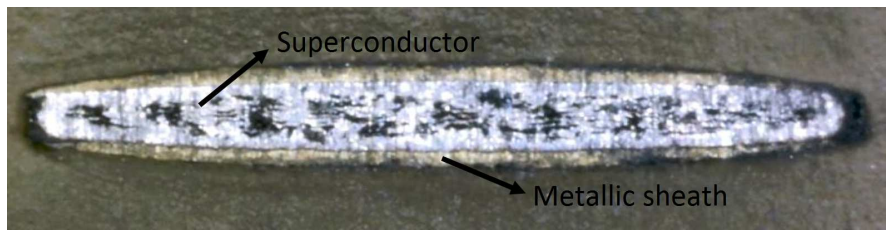

FIG. 2: Micrograph of the cross-section of the tape.

### III. RESULTS AND DISCUSSION

In Fig.3 the $\langle E \rangle - J$ dependencies for longitudinal and transverse bridges for different values of temperatures are shown. Panel (a) shows the dependence of the current density with the electric field for the longitudinal bridge and panel (b) for the transverse one. We let $\langle E \rangle$ denote the average field between a pair of voltage contacts separated by a distance $l$, and compute $V/l$ to determine it from the measured voltage drop, $V$. The current density $J$ was obtained from the measured current $I$ through the equation $J = I/A$, where $A$ is the area of the cross-section of the bridges.

It can be seen from Fig.3(a) that the dependence obtained for the lowest temperature (circles,100K) shows no substantial dissipation in the measured range. Nevertheless, this curve has a small slope of unknown source. The curve for 110K (triangles,110K) is still in the superconducting state and has a critical current density of $3.20 \pm 0.32$ $A/mm^2$. The electric field criterion used here was $\langle E \rangle_c = 0.02$ V/m. Notice that after the current density reaches its critical value, the two slopes corresponding to the flux-flow and normal state can be clearly identified. The dependence for 120K (squares,120K) shows an ohmic behavior as expected, because at that temperature the sample is in the normal state.

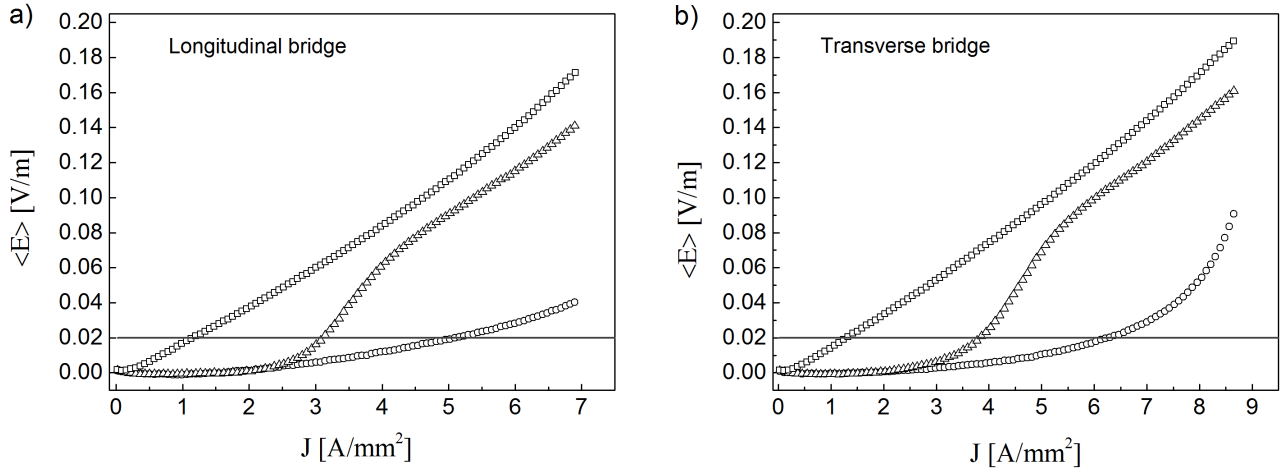

FIG. 3: $\langle E \rangle - J$ curves for different values of temperature. (a) $\langle E \rangle - J$ dependence for the longitudinal bridge and (b) for the transverse bridge. The curves for 100K are represented by circles, the triangles represent the curves for 110K, and the curves for 120K are represented by squares.

The $\langle E \rangle - J$ curves in Fig.3(b) corresponding to the transverse bridge show a similar qualitative behavior but not a quantitative one, when compared to those obtained for the longitudinal bridge. It can be seen from Fig.3(b) that the curve corresponding to 100K (circles) exhibit a critical current density of $6.00 \pm 0.60$ $A/mm^2$ and the curve corresponding to 110K (triangles) shows a $J_c$ of $3.80 \pm 0.38$ $A/mm^2$. The latter value of critical current density for the transverse bridge is substantially higher than its analogous for the longitudinal bridge. This result is unexpected: the superconducting properties of the system are anisotropic in spite of the fact that the sample is not multi-filamentary. Another unexpected result is that the critical current density is substantially larger in the transverse direction than in the longitudinal one. The typical situation observed in multi-filamentary tapes is quite different [26–28].

The anisotropy found can be quantified using the lateral-transverse anisotropy coefficient [28]:

$$A_{lt} = 1 - \frac{E_{long}}{E_{trans}} \quad (1)$$

where $E_{long}$ stands for the electric field measured in the longitudinal direction and $E_{trans}$ represents the electric field in the transverse direction. It is worth pointing out that the $A_{lt}$ coefficient depends on the electric current and also depends on the temperature. In the case of multi-filamentary BSCCO tapes, this coefficient is positive and goes from zero (when the dissipation is the same in both directions) to one (when the longitudinal bridge behaves as a perfect superconducting path to the current and the dissipation in this direction is zero). Applying equation 1 to the electric field values extracted from Fig.3 it can be seen that the anisotropy coefficient has negative values due to the unusual anisotropy presented on the tape where the transverse direction shows better transport properties than the main direction of the tape. This anomalous anisotropy might be caused by the effects of the substrate in the microstructure of the superconductive phase producing a "stronger" superconductivity in the transverse direction.

## IV. CONCLUSION

We have been able to directly measure the transverse transport properties of a BSCCO-Pb high temperature superconducting tape, showing that there is an anisotropy between this direction and the main direction of the tape in spite of the fact that the tape is not multi-filamentary.

It was found that the critical current density is substantially larger in the transverse direction of the tape than in the longitudinal one, then showing an anomalous anisotropy. This behavior might be caused by the effects of the internal structure of the substrate in the growing of the superconducting phase. The study of this anomalous anisotropy is potentially relevant for AC applications, and also in a situation of failure due to transversal cracks, where currents are forced to flow perpendicular to the long direction of the tape.